\documentstyle[preprint,prabib,aps,12pt]{revtex}

\begin{document}
%\draft
\title{4D Wormhole with  Signature Change in the 
Presence of  Extra Dimensions}
\author{V. Dzhunushaliev
\thanks{E-Mail Addresses : dzhun@rz.uni-potsdam.de and 
dzhun@freenet.bishkek.su; \qquad permanent address: 
Dept. Theor. Phys., Kyrgyz State National University, 
Bishkek 720024, Kyrgyzstan}}
\address{Institut f\"ur Mathematik, Universit\"at Potsdam 
PF 601553, D-14415 Potsdam, Germany} 
\author{H.-J. Schmidt
\thanks{http://www.physik.fu-berlin.de/\~{}hjschmi \ \ 
 \quad  hjschmi@rz.uni-potsdam.de}}
\address{Institut f\"ur Theoretische Physik, Freie
Universit\"at Berlin\\
and \\
Institut f\"ur Mathematik, Universit\"at Potsdam 
PF 601553, D-14415 Potsdam, Germany} 

\maketitle

\begin{abstract}
A  regular vacuum  solution in  5D gravity on the principal bundle  
with the $U(1)$ structural group is proposed as a 4D wormhole. 
This solution has two null  hypersurfaces where an interchange of the sign 
of some 5D metric components happens. For a 4D observer living on the  
base of this principal bundle this is a wormhole with two 
asymptotically flat Lorentzian (Euclidean) spacetimes connected 
by a  Euclidean (Lorentzian) throat. The 4D Lorentzian observer sees these 
two null  hypersurfaces as  electric charges. 
\end{abstract}

\section{Introduction}

The nice Hawking idea about a change of the signature of the spacetime 
metric has a problem in the classical regime: usually, 
 a singularity  appears 
at  that  point where this change takes place. The simplest
 explanation for this 
is the following: the determinant
 of the metric tensor $g = \det (g_{ik})$ changes 
its sign by changing the metric signature. Therefore at  this 
point $g = 0$ and/or one of the 
 scalars $R$ or $R_{ik}R^{ik}$ or $R_{iklm}R^{iklm}$ 
is equal to $\pm \infty$. A detailed  explanation of this fact can be 
found in 
\cite{ellis97} and the bibliography for this subject. 
\par 
It can also be shown  that the  gravitational field requires 
additional 
degrees of freedom for the change of metric signature. It is easy 
to see if we write the metric in the vier-bein formalism: 
\begin{equation}
ds^2 = \eta_{ab} \omega ^a \omega ^b ,
\label{1}
\end{equation}
here $\omega ^a = e^a_\mu dx^\mu$; $\eta _{ab} = (+1,-1,-1,-1)$ 
is the Minkowski metric; $e^a_\mu$ is the  
vier-bein.\footnote{$e^a_\mu$ are the
 degrees of freedom of the gravitational 
field, this means that the gravitational equations are deduced 
by varying with respect to vier-bein $e^a_\mu$.} The signature of 
the metric is defined by $\eta _{ab}$ and is not varying. It 
 is possible that the 
 change
 of the metric signature can occur as a quantum 
process on the spacetime foam level when $\eta _{ab}$ is changed. 
\par 
But below we will show that in the 5D Kaluza-Klein gravity there 
is a trick with interchanging of the sign between some 5D metric 
components 
that for the 4D observer is similar 
 to the change of the signature of the 4D metric. 

\section{Signature change in the 4D wormhole}

In \cite{dzh16}, \cite{dzh2}  the following 
wormhole-like (WH) solution 
in the vacuum 5D Kaluza-Klein gravity was found: 
\begin{equation}
ds^{2}_{(5)} = - \frac{r_0^2}{\Delta (r)}(d\chi  - \omega (r)dt)^2 + 
\Delta (r)dt^{2} - dr^{2} - a(r)d\Omega ^2 ,
\label{2}
\end{equation}
here $\chi $ is the 5$^{th}$ extra coordinate; 
$r,\theta ,\varphi $ are the $3D$  polar coordinates; 
$t$ is the time; 
$d\Omega ^2 = d\theta ^{2} + \sin ^{2}\theta  d\varphi ^2$ is the 
metric on the $S^2$ sphere; the subscript  (5)  
 denotes that the appropriate quantity is 5 dimensional. 
The equations for $\Delta (r)$ are: 
\begin{eqnarray}
\frac{\Delta ''}{\Delta} - \frac{{\Delta '}^2}{\Delta ^2} +  
\frac{a' \Delta '}{a\Delta} - \frac{r_0^2}{\Delta ^2}{\omega '}^2 & = & 0 ,
\label{2-1}\\
\omega '' - 2 \omega ' \frac{\Delta '}{\Delta} + 
\omega ' \frac{a'}{a} & = & 0 ,
\label{2-2}\\
\frac{{\Delta '}^2}{\Delta ^2} + \frac{4}{a} - 
\frac{{a'}^2}{a^2} - \frac{r_0^2}{\Delta ^2}{\omega '}^2 & = & 0 ,
\label{2-3}\\
a'' - 2& = & 0 .
\label{2-4}
\end{eqnarray}
and we see that if $\Delta$ is a solution 
then $-\Delta$ is also.\footnote{In contrast to this example, 
for the Schwarzschild black hole this is not the case.  
For the metric $ds^2 = \Delta dt^2 - dr^2/\Delta -r^2d\Omega ^2$ 
there we get the following equation: 
$\Delta ' + \Delta /r -1/r =0$ which is not invariant under 
$\Delta \rightarrow - \Delta$ transformation.}. 
The solution of these $5D$ Einstein's equations is
\begin{eqnarray}
a & = & r^{2}_{0} + r^{2},
\label{3}\\
\Delta & = & \pm \frac{2r_0}{q}\frac{r^2 + r_0^2}
{r^2 - r_0^2} ,
\label{4}\\
\omega & = &  \pm \frac{2r_0^2}{q}\frac{a'/a}
{1 - \frac{2r_0^2}{a}} .
\label{5}
\end{eqnarray}
(i.e., $\omega=2r r_0 \Delta/a$), 
here $r_0 > 0$ and $q$ are the same constants. 
\par 
In this paper the 5D spacetime is the total space of the principal bundle 
with the $U(1)$ group as the structural group, the base of this bundle is 
the ordinary 4D spacetime \cite{dzh2}. This means that we condider the
following part of metric (2):
$$
ds_{(4)}^2 = \Delta (r)dt^{2} - dr^{2} - a(r)d\Omega ^2
$$
 For the metric on the total 
space $E$ of the 
principal bundle the most natural choice of the coordinate system  
is the following: the $y^a$ ($a = 5,6, \cdots$) 
coordinates\footnote{In our case, we restrict to the 
1-dimensional fibre, i.e. $y^5 = \chi$, and the index $a=5$.}  
 are chosen on the fibre 
(gauge group) and $x^\mu$ ($\mu = 0,1,2,3$) 
along the base (4D spacetime). In this case 
$y^a$ are the coordinates which cover the gauge group $G$ 
(the fibre of the bundle) and 
$x^\mu$ are the coordinates which cover the factor space $E/G$ 
(the base of the bundle or 4D spacetime). 
\par 
In the classical and quantum field theories (without gravitation) 
the strong, weak and electromagnetic interactions are characterized as 
a connection on the appropriate bundle with some structural group. 
These fields are real,  therefore the corresponding total space is also real 
in  Nature. But we cannot choose the new coordinate system 
in which we will mix the coordinates on the fibre (gauge group) 
and the base (4D spacetime). This is evident  because the 
points on the fibre are the elements of the group but the points 
on the base are  not. In the 
context of this paper the metrc  (\ref{2}) %metric
 is the metric on the 
total space.
Once again we emphasize that the Kaluza-Klein 
theory in the  context of this paper is the gravity on such 
a principal bundle, therefore we can use only the following  coordinate 
transformations: 
\begin{eqnarray}
y'^a & = & y'^a (y^a) + f^a\left (x^\mu \right ) ,
\label{6}\\
x'^\mu & = & x'^\mu \left (x^\mu \right ) .
\label{7}
\end{eqnarray}
The first term in (\ref{6}) means that the choice of coordinate 
system on the fibre is arbitrary. The second term indicates that 
in addition we can move the origin of coordinate system on each 
fibre on the value $f^a(x^\mu)$. It is well known that such a 
transformation law   (\ref{6})  leads to a local gauge transformation for 
the appropriate nonabelian field (see for overview \cite{overduin:1997pn}). 
That is the (\ref{6}) and (\ref{7}) coordinate transformation 
are the most {\it natural} transformation for the multidimensional 
gravitation on the principal bundle. Of course we can use the 
much more generalized coordinate transformations: 
\begin{eqnarray}
y'^a & = & y'^a \left (y^a, x^\mu \right ) ,
\label{8}\\
x'^\mu & = & x'^\mu \left (y^a, x^\mu \right ) .
\label{9}
\end{eqnarray}
But in this case we destroy the initial topological structure 
of the multidimensional spacetime, and what is even worse
 we mix the 
points of the fibre 
\footnote{which are the elements of some group.} 
with the points of the base 
\footnote{which are the ordinary spacetimes points.}. 
That we do not  in  ordinary classical/quantum field theory 
(without gravitation) hence this would be  a bad coordinate choice 
for the multidimensional gravity on the principal bundle. 
\par 
The above-mentioned item is a literary description for the next 
exact theorem \cite{per1,sal1}: 
\par
Let $G$ be the group 
fibre of the principal  bundle.  Then  there  is a one-to-one
correspondence between the $G$-invariant metrics on the  total  
space ${\cal X}$: 
\begin{equation}
ds^2 = G_{AB} dx^A dx^B = 
g_{\mu\nu} + h(x^{\mu}) \left ( \omega ^a + 
A^a_\mu dx^\mu \right )^2 
\end{equation}
and the triples $(g_{\mu \nu }, A^{a}_{\mu }, h)$. 
Here $G_{AB}$ is the multidimensional metric on the total space 
($A,B = 0,1,2,3,5,6, \cdots$)
$g_{\mu \nu }$ is Einstein's pseudo  -
Riemannian metric on the base; $A^{a}_{\mu }$ is the gauge field 
of the group $G$ ( the nondiagonal components of 
the multidimensional metric); $h\gamma _{ab}$  is the 
symmetric metric on the fibre; $\omega _a = \gamma _{ab}\omega ^b$; 
$\omega ^a$ are the one-form on the group $G$. 
\par 
In  Ref.\cite{dzh7} the solution (\ref{3}-\ref{5}) %solution
 was applied 
for the discussion %reception 
of the composite Lorentzian WH. For this goal the WH-like 
solution (\ref{2}) (with 
$|r| \leq r_0$, and the sign (-) in (\ref{4}), (\ref{5})) 
is inserted between two Reissner-Nordst\"om black holes. 
\par 
Below we examine two possibilities for the signs (+) and (-) 
in eqs. (\ref{4}), (\ref{5}). 

\subsection{Lorentzian wormhole with the Euclidean throat, the case 
of (+)}
\label{lorentz}

Here we examine the solution (\ref{2}) with (7,8,9)  in the whole  region 
$-\infty < r < +\infty$. In this case, for the 4D spacetime (the base of 
the principal bundle) the following  takes place: 
\begin{enumerate}
\item
By $|r| \ge r_0$ we have the ordinary 4D asymptotically flat 
spacetime with Lorentzian signature (from the  viewpoint of  the 
4D observer) 
as $g_{tt} = \Delta > 0$.
\item
By $|r| < r_0$ we have the Euclidean 4D spacetime bounded between 
two $ds^2_{(5)} = 0$ (located by $r = \pm r_0$) hypersurfaces as 
$g_{tt} = \Delta < 0$.
\item 
From the viewpoint of 4D the observer on the $r = \pm r_0$ hypersurfaces 
takes place the change of the 4D metric signature. This is a result 
of simple interchange the signs of the metric components $G_{tt}$ 
and $G_{55}$ on the $ds^2_{(5)} = 0$ hypersurfaces and nothing  more.
\end{enumerate}
Thus, the solution (\ref{2})  describes the WH connecting two Lorentzian 
asymptotically flat regions by means of the Euclidean throat. 
\par 
Of course we have a question: what happens at the
 hypersurfaces  $r = \pm r_0$?
 Now we describe the properties of such hypersurface 
on which the interchange of the metric signature happens: 
\begin{enumerate}
\item
On this surface $ds^2_{(5)} = 0$ ($\chi$, $\theta$, $\varphi =const$ 
and $r = \pm r_0$). 
\item 
\label{nonsingular}
The 5D scalar invariants $R_{(5)} = R_{(5)AB}R_{(5)}^{AB} = 0$ 
in the consequence 
of the 5D Einstein equations $R_{(5)AB} = 0$. 
$R_{(5)ABCD}R_{(5)}^{ABCD} \propto r_0^{-4}$ ($A=0,1,2,3,5$), 
i.e. we see that probably these two 
$ds^2_{(5)} = 0$ hypersurfaces do not have a singularity. 
\item 
The 4D metric on the 4D base of the  principal bundle on the 4D spacetime
 is  (for $a$ see eq. (7)):  
\begin{equation}
ds^2_{(4)} = {2r_{0}\over q}{1\over 1 - {2r^{2}_{0} \over a}} dt^2 - 
dr^2 - \left ( r^2 + r_0^2\right )d\Omega ^2 .
\label{6a}
\end{equation}
By $r \rightarrow \pm \infty$ we have two asymptotically flat Lorentzian 
spaces with the metric: 
\begin{equation}
ds^2_{(4)} \approx \frac{2r_0}{q}
 dt^2 - dr^2 - r^2 d\Omega ^2
\label{7a}
\end{equation}
\item 
The 4D curvature scalar $R_{(4)}$: 
\begin{equation}
R_{(4)} = \frac{6r_0^2}{\left (r^2 - r_0^2\right )^2} ,
\label{8a}
\end{equation}
and it has the singularity. Thus, from the point of view of 4D the observer 
there is a singularity. But we emphasize once again that this 
singularity is not the really singularity in the consequence of the 
second  item. This situation is similar to what happens 
in the Schwarzschild metric: 
\begin{equation}
ds^2_{(4)} = \left (1 - \frac{r_g}{r} \right ) dt^2 - 
\frac{dr ^2}{1 - \frac{r_g}{r}} - r^2 d\Omega ^2
\label{9a}
\end{equation}
on the event horizon. 
\item 
For the 4D observer in the Lorentzian part of this WH these two 
singularities look as two $(\pm)$ electric charges spreaded on the 
$r = \pm r_0$ surfaces 
with the outgoing and 
incoming force lines of the electric field. 
\item 
In this 5D case we cannot introduce the notion of electric 
charge. To see it more directly we shall look on the eq. (\ref{2-2}), 
  it can be  rewritten   in the following form: 
\begin{equation}
\left (4\pi a \frac{\omega '}{\Delta ^2} \right )' = 0 .
\label{10}
\end{equation}
This means that the product of the electric field 
$F_{01} = \omega '$ on the area   $4\pi a$  of the sphere $S^2$
is not a conserved electric charge $q$. But interesting is 
that we can correct this field: as we see from the (\ref{10}) 
the product of magnitude 
$\omega '/\Delta ^2$ with the area   $4\pi a$  of the sphere $S^2$ %sphere $4\pi a$ 
is the conserved flux of the corrected electric field which 
is proportional to the electric charge. 
\end{enumerate} 
It is interesting to compare the metric (\ref{2})  with the 
Reissner-Nordstr\"om solution. For this purpose we introduce 
the new radial coordinate $\rho = \sqrt{r^2 + r_0^2}$. Then we have 
our metric in the following form: 
\begin{equation}
ds^2=- \frac{r_0q}{2} \left (1 - \frac{2r_0^2}{\rho ^2} \right ) 
\left (d\chi - \omega dt\right )^2 + \frac{2r_0}{q} 
\frac{dt^2}{1 - \frac{2r_0^2}{\rho ^2}} 
- \frac{d\rho ^2}{1 - \frac{r_0^2}{\rho ^2}} - 
d\Omega ^2 .
\label{11}
\end{equation}
here the area of the sphere $S^2$  is $4\pi \rho ^2$ as for the 4D 
Reissner-Nordstr\"om solution but 
$g_{tt} = \frac{2r_0}{q} \left (1 - \frac{2r_0^2}{\rho ^2}\right )^{-1}$ 
and 
$g_{\rho \rho } = \left (1 - \frac{r_0^2}{\rho ^2} \right )^{-1}$ 
differ from the corresponding metric components of the 
Reissner-Nordstr\"om solution. Also the Maxwell tensor for the 
 metric (\ref{2}) is 
$F_{01} = \omega ' = \frac{\Delta ^2}{r_0}\frac{q}{\rho ^2}$ 
whereas for the Reissner-Nordstr\"om metric we have 
$F_{01} = E = \frac{q}{r^2}$. 
Hence we can say that the metric (\ref{2})  cannot be 
considered as the model of 4D ``charge without charge''. 
This is a simple example of a possible signature change in the 
presence of the extra dimensions. 

\subsection{Euclidean wormhole with the Lorentzian throat, 
the case (-)}

Here we can precisely repeat our reasoning of subsection  \ref{lorentz} 
 with the following interchanging: $Euclidean 
{\longrightarrow \atop \longleftarrow} Lorentzian$. 
Thereof we have the WH with the Lorentzian throat connecting two 
Euclidean 
asymptotically flat regions. 

\section{Conclusion}

The basic idea of the Kaluza-Klein paradigm is that the extra dimensions 
are very small and therefore unobservable. If so then a  wormhole with 
metric (\ref{2}) can be a simple example of 
an Euclidean bridge between two Lorentzian regions 
(or the Lorentzian bridge between two Euclidean regions). 
We remark that this 5D construction is regular everywhere, i.e. 
there is not any singularity in this solution. It is remarkable 
that this solution is a regular {\it vacuum}  solution for the 5D 
Kaluza-Klein gravity in the spirit of Einstein's  idea that the 
right-hand 
 side of the gravitational field  equations should be \underline{zero}. 
\par  
Finally, we can say that by the assumption of the hidden extra dimensions 
in the Kaluza-Klein theory there is a possibility for the 
signature change of the 4D metric.

\bibliography{john,john2}

\begin{thebibliography}{1}

\bibitem{ellis97}
C. Hellaby, A. Sumeruk, and G.~F.~R. Ellis, Int. J. Mod. Phys. {\bf D6},  211
  (1997); R. Mansouri et al., Gen. Relat. Grav. in print; 
 G. Ellis et al. Gen. Relat. Grav. {\bf 29}, 591 (1997).

\bibitem{dzh16}
V. Dzhunushaliev, Izv. Vuzov, ser. ''Fizika``  78  (1993).

\bibitem{dzh2}
V. Dzhunushaliev, Gen. Relat. Grav. {\bf 30},  583  (1998).

\bibitem{overduin:1997pn}
J.~M. Overduin and P.~S. Wesson, Phys. Rept. {\bf 283},  303  (1997).

\bibitem{per1}
R. Percacci, J. Math. Phys. {\bf 24},  807  (1983).

\bibitem{sal1}
A. Salam and J. Strathdee, Ann. Phys. {\bf 141},  316  (1982).

\bibitem{dzh7}
V. Dzhunushaliev, Mod. Phys. Lett. A {\bf 13},  2179  (1998).

\end{thebibliography}
\bibliographystyle{prsty}

\end{document}